\documentclass[aps,prl,reprint,floatfix]{revtex4-2}

\usepackage{graphicx}
\usepackage{tikz}
\usepackage{hyperref}
\usepackage{physics}
\usepackage{amsfonts,amsmath}

\begin{document}

\title{Efficient tensor network algorithm for layered systems}

\author{Patrick C.G. Vlaar}
\email[]{p.c.g.vlaar@uva.nl}
\affiliation{Institute for Theoretical Physics and Delta Institute for Theoretical Physics,
             University of Amsterdam, Science Park 904, 1098 XH Amsterdam, The Netherlands}

\author{Philippe Corboz}
\affiliation{Institute for Theoretical Physics and Delta Institute for Theoretical Physics,
             University of Amsterdam, Science Park 904, 1098 XH Amsterdam, The Netherlands}

\date{\today}

\begin{abstract}
Strongly correlated layered 2D systems are of central importance in condensed matter physics, but their numerical study is very challenging. Motivated by the enormous successes of tensor networks for 1D and 2D systems, we develop an efficient tensor network approach based on infinite projected entangled-pair states (iPEPS) for layered 2D systems. Starting from an anisotropic 3D iPEPS ansatz, we propose a contraction scheme in which the weakly-interacting layers are effectively decoupled away from the center of the layers, such that they can be efficiently contracted using 2D contraction methods while keeping the center of the layers connected in order to capture the most relevant interlayer correlations. We present benchmark data for the anisotropic 3D Heisenberg model on a cubic lattice, which shows close agreement with quantum Monte Carlo (QMC) and full 3D contraction results. Finally, we study the dimer to N\'eel phase transition in the Shastry-Sutherland model with interlayer coupling, a frustrated spin model which is out of reach of QMC due to the negative sign problem.
\end{abstract}

\maketitle
Understanding the emergent phenomena in strongly correlated systems is of central importance in modern physics. Among the most powerful tools to study these systems are tensor network (TN) methods, with the density matrix renormalization group (DMRG)~\cite{White1992} algorithm and its underlying variational ansatz, the matrix product state (MPS)~\cite{Fannes1992,Ostlund1995}, being the best-known examples for (quasi) one-dimensional systems. Projected entangled-pair states (PEPS)~\cite{Verstraete2004,Murg2007} (or tensor product states~\cite{Hieida1999,Maeshima2001,Nishio2004}) provide a natural generalization of MPS to higher dimensions. Thanks to algorithmic advances in the past years, PEPS has become a versatile state-of-the-art tool for 2D systems, not only for ground states~\cite{corboz14_shastry, corboz14_tJ, liao16, zheng17, niesen17, chen18, jahromi18, lee18, yamaguchi18, ponsioen19, kshetrimayum20, chung19, haghshenas19, Chen20, lee20, gauthe20, hasik21, liu21}, but  also for finite temperature calculations~\cite{Czarnik2017, Peng2017, Chen2018, kshetrimayum19, Czarnik2019, wietek19, czarnik19c, czarnik21, jimenez21, poilblanc21, gauthe22},  excited states~\cite{Vanderstraeten2019,Ponsioen2020,ponsioen22,chi22}, open systems~\cite{Kshetrimayum2017,Czarnik2019,kilda21,mckeever21}, and real-time evolution~\cite{Czarnik2019,Hubig2019,Hubig2020,Kshetrimayum2020,kshetrimayum21,schmitt21,dziarmaga22b}. 3D TN algorithms are more challenging because of their higher complexity, although progress has recently been made in developing methods with a tractable computational cost~\cite{Vlaar2021} (see also related works on 3D classical systems~\cite{nishino1998,nishino2000,nishino2001,xie12,orus12,vanderstraeten18}).

A special and highly relevant class of 3D quantum systems is formed by layered 2D systems, in which the effective intralayer couplings are much stronger than the interlayer ones. Important realizations include the cuprate high-T$_c$ superconductors~\cite{bednorz86} as well as various quasi-2D frustrated magnets such as Kagom\'e~\cite{shores05}, triangular~\cite{coldea01, shimizu03, Shirata2012, rawl17, cui18}, Shastry-Sutherland~\cite{Kageyama99,Miyahara99,Miyahara2003}, and honeycomb lattice compounds~\cite{singh10,plumb14,takagi19,wessler20}. While pure 2D models often already capture the relevant physics of these systems, the interlayer couplings can play an important role on the quantitative level. For example, they lead to a finite N\'eel transition temperature in layered square lattice Heisenberg models as opposed to the pure 2D case~\cite{Sengupta2003}, or they may play a significant role in the competition of low-energy states in the 2D Hubbard model~\cite{zheng17}. Thus, accurate TN approaches to study these systems would be highly desirable.

In this letter, we introduce an efficient TN algorithm for  layered 2D systems, called the layered corner transfer matrix (LCTM) method, which is substantially simpler and computationally cheaper than full 3D approaches. Motivated by the layered nature of these systems, we start from an anisotropic PEPS ansatz, i.e., with a small interlayer bond dimension $D_z$ compared to the intralayer bond dimension $D_{xy}$, which control the accuracy of the ansatz. The main idea of the algorithm is to contract the 3D TN by (1) performing an effective decoupling of the layers away from the center of each layer, (2) contracting the individual decoupled layers using the standard 2D corner transfer matrix (CTM) method~\cite{Nishino1996,Orus2009,corboz14_tJ}, and (3) contracting the remaining TN, formed by the contracted layers connected with a finite bond dimension $D_z>1$ in the center of each layer. A core ingredient of the approach is the effective decoupling procedure which we implement based on an iterative full-update (FU) truncation scheme~\cite{Jordan2008,Phien2015}. We present benchmark results for the anisotropic Heisenberg model on a cubic lattice, which show close agreement with a full 3D contraction and with quantum Monte Carlo (QMC) data already for small $D_z$. As a more challenging example we consider a frustrated spin model, the Shastry-Sutherland model with interlayer coupling, for which QMC fails due to the negative sign problem. Finally, we highlight directions for future improvements and extensions of the LCTM approach.

\begin{figure}
    \centering
    \includegraphics[width=1\linewidth]{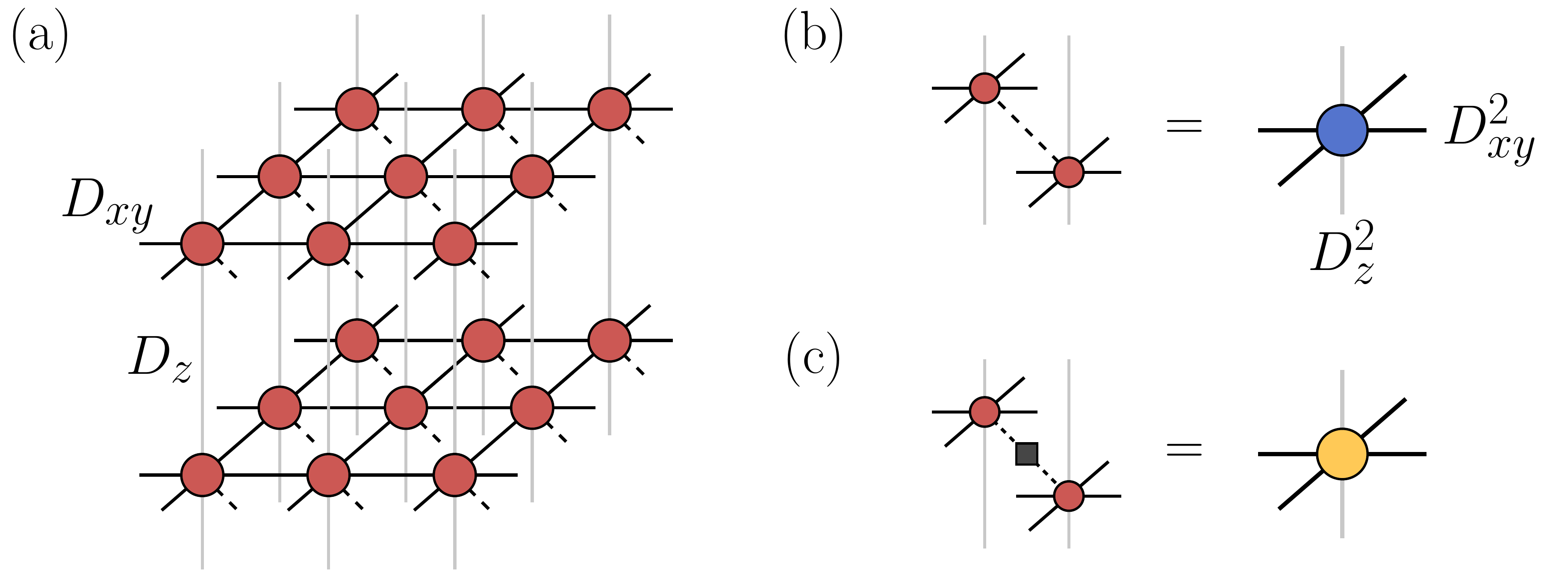}
    \caption{(a) Anisotropic 3D iPEPS ansatz with intralayer and interlayer bond dimensions $D_{xy}$ and $D_{z}$, respectively. 
    (b) The norm tensor represents the combined bra- and ket-tensors on each site where pairs of auxiliary indices are combined into new auxiliary indices with dimensions $D_{xy}^2$ and $D_{z}^2$, respectively.  (c) Same as in (b) but with a local operator between the bra- and ket-tensors  for the evaluation of a local expectation value.}
    \label{fig:ipeps_layered}
\end{figure}

\emph{Method.} --- 
We consider an infinite PEPS (iPEPS), shown in Fig.~\ref{fig:ipeps_layered}(a), consisting of a tensor (or more generally a unit cell of tensors) that is repeated on the infinite cubic lattice. Each tensor has 7 indices: one physical index carrying the local Hilbert space of a site, four indices with bond dimension $D_{xy}$ connecting to the intraplane nearest-neighbor tensors, and two indices of dimension $D_z$ connecting to the tensors in the neighboring planes. The accuracy of the ansatz is systematically controlled by $D_{xy}$ and $D_z$, where we choose $D_{xy}\geq D_z$ motivated by the anisotropic nature of layered 2D systems. In the limit of $D_z=1$, the ansatz corresponds to a product state of 2D iPEPS layers, i.e., a state without  entanglement between the layers (but with entanglement within the layers, controlled by $D_{xy}$).
\begin{figure}
    \centering
    \includegraphics[width=\linewidth]{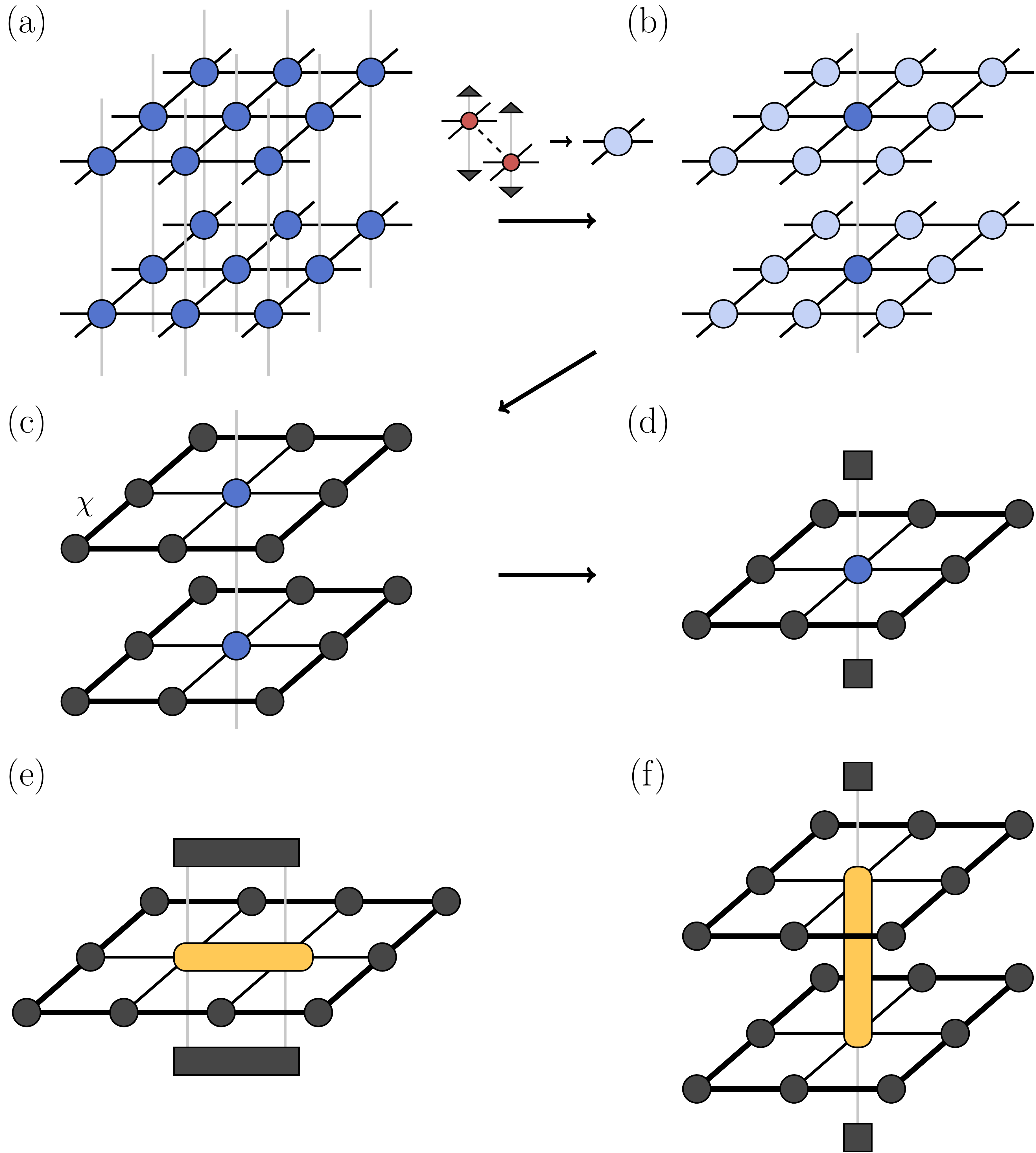}
    \caption{The main steps of the LCTM contraction method. (a)~3D TN representing the norm. (b)~TN with decoupled layers away from the center, obtained by a projection (black triangles) of the vertical indices of bra- and ket-tensors in (a) onto $D_z=1$ (except in the center). (c) The decoupled 2D layers are contracted using the CTM method, which yields the environment tensors around the central tensor, i.e., four corner and four edge tensors (in black), with an accuracy controlled by the boundary bond dimension $\chi$. (d)~The infinite central chain with contracted layers can be evaluated by replacing the neighboring layers by the left- and right-dominant eigenvector (black squares) of the transfer matrix represented by a contracted layer. A local expectation value can be computed by replacing the central norm tensor by the yellow tensor shown in Fig.~\ref{fig:ipeps_layered}(c). (e)-(f) Relevant diagrams to compute  intra- and interlayer nearest-neighbor observables, where the yellow tensor is obtained from contracting neighboring bra- and ket-tensors  with a 2-site operator in between.}
    \label{fig:layered_ctm_algorithm}
\end{figure}

The main challenge of a 3D TN algorithm is the efficient, approximate contraction of the 3D TN, which is needed to compute, e.g., a local expectation value. Let us consider computing the norm of the wave function. The corresponding TN is depicted in Fig.~\ref{fig:layered_ctm_algorithm}(a), where the norm tensors (blue) represent the combined bra- and ket-tensors on each site as shown in Fig.~\ref{fig:ipeps_layered}(b). To compute a local expectation value, we can simply put an operator between the local tensors as shown in Fig.~\ref{fig:ipeps_layered}(c) and replace the norm tensor with this new tensor at the desired location. 

In the simplest case, for $D_z=1$, this network consists of independent 2D square lattice networks which can be efficiently contracted using the CTM method~\cite{Nishino1996,Orus2009,corboz14_tJ}. The CTM method is an iterative approach that approximates the 2D TN surrounding a central tensor by a set of environment tensors, given by four corner and four edge tensors (shown by the black disc-shaped tensors in Fig.\ref{fig:layered_ctm_algorithm}(c)), where the accuracy is systematically controlled by the boundary bond dimension $\chi$ of the environment tensors.

For the case $D_z>1$, a full 3D contraction algorithm as in Ref.~\cite{Vlaar2021} could be used. This, however, is computationally expensive, and we thus follow a more efficient strategy here, exploiting the anisotropic nature of the ansatz. The main idea is to project the vertical indices of the bra- and ket-iPEPS tensors away from the center of each layer onto $D_z=1$ (see details below), while keeping the full bond dimension $D_z>1$ on the tensors in the center, see  Fig.~\ref{fig:layered_ctm_algorithm}(b). This leads to an effective decoupling of the 2D layers away from the center, such that the standard 2D CTM approach can be used to contract them (Fig.~\ref{fig:layered_ctm_algorithm}(c)) while the most relevant interlayer correlations are still taken into account by the vertical connections of the tensors in the center. Since the bonds in the z-direction carry only little entanglement, the projection onto $D_z=1$ away from the center is expected to induce only a small error on a local expectation value measured in the center. After contracting each layer, the resulting TN corresponds to an infinite 1D chain in the vertical direction with bond dimension $D_z^2$, which can be evaluated by sandwiching the central layer between the left- and right-dominant eigenvector of the corresponding transfer matrix (represented by a contracted layer), as shown in Fig.~\ref{fig:layered_ctm_algorithm}(d).

A core ingredient of the algorithm is the projection step from $D_z>1$ to $D_z=1$. We use a scheme based on a full-update (FU) truncation~\cite{Jordan2008,Phien2015}, a technique that is also applied in the context of imaginary time evolution algorithms to truncate a bond index in an iPEPS. It is based on a minimization of the norm distance, $ d =  \begin{Vmatrix} \ket{\psi} - \ket{\psi'}  \end{Vmatrix}^2 $, which can be solved iteratively, where $\ket{\psi}$ is the untruncated iPEPS and $\ket{\psi'}$ is the iPEPS with one bond truncated down to $D_z=1$, see the supplemental material~\cite{suppl_mat} for details. The FU does, however, require the environment tensors, which we initially do not have. We thus start from an initial approximate projection based on the simple update (SU) approach~\cite{Jiang2008}, which only considers local tensors for the truncation, from which the environment for the FU projection is computed. To improve the accuracy of the truncation, one can repeat the computation of the environment iteratively. In practice, for the model considered here, we find that one FU iteration is sufficient to reach convergence. 

The accuracy of the LCTM method is controlled by the boundary bond dimension $\chi$ and by the number of $D_z>1$ connections kept in the center. Here, we focus on the simplest case, where we only keep the connections on the central tensor for the evaluation of one-site observables and interplane two-site observables (see Fig.~\ref{fig:layered_ctm_algorithm}(f)), which we find is sufficient in the limit of weak interlayer coupling, as we will show in our benchmark results. For intralayer two-site observables we keep two connections, as depicted in Fig.~\ref{fig:layered_ctm_algorithm}(e). The computational cost of these contractions are $\chi^3 D_{xy}^4 + \chi^2 D_{xy}^6 D_z^2$ and $\min [\chi^3 D_{xy}^6, \chi^3 D_{xy}^4 D_z^4]$, respectively. In the supplemental material~\cite{suppl_mat} we discuss other layer decoupling approaches and we also consider a scheme with more connections, which is more accurate, but also computationally more expensive.

The LCTM contraction method can not only be used for the computation of observables, but also in combination with accurate optimization schemes (to find the optimal variational parameters in the tensors for a given Hamiltonian), e.g., in an imaginary time evolution with fast-full update (FFU)~\cite{Phien2015} or in energy minimization algorithms~\cite{corboz16b,vanderstraeten16,liao2019}. We further note that the LCTM method can be extended to arbitrary unit cell sizes in a similar way to the standard CTM in 2D~\cite{corboz2011,corboz14_tJ}.

\emph{Results.} ---  
To benchmark the method, we consider the anisotropic 3D Heisenberg model on a cubic lattice given by the Hamiltonian
\begin{equation}
    \hat{H} = J_{xy} \sum_{\expval{ij}_{xy}} \mathbf{S}_i \mathbf{S}_j + J_z \sum_{\expval{ij}_z} \mathbf{S}_i \mathbf{S}_j,
\end{equation}
with $J_{xy}$ the intralayer and $J_z$ the interlayer coupling strengths and $ \mathbf{S}_i$ spin $S=1/2$ operators. We use an iPEPS ansatz with two tensors, one for each sublattice, to capture the long-range antiferromagnetic order. The iPEPS is optimized with the FFU imaginary time evolution algorithm~\cite{Phien2015}, starting from initial tensors obtained with simple update optimization~\cite{Jiang2008}. In the CTM approach, we keep a sufficiently large boundary bond dimension $\chi$, such that finite-$\chi$ effects are negligible~\cite{suppl_mat}. 
To improve the computational efficiency, tensors with implemented $U(1)$ symmetry~\cite{Singh2011,Bauer2011} are used~\footnote{We note that the SU(2) spin symmetry is broken in the ground state.}. We compare our results to the ones computed with the full 3D contraction approach (SU+CTM) from Ref.~\cite{Vlaar2021} and with QMC results based on the directed loop algorithm from the ALPS library ~\cite{Albuquerque2007,Bauer2011(2)} (obtained at a sufficiently low temperature  of $T=0.005 J_{xy}$). To extrapolate the QMC data to the thermodynamic limit, a finite size scaling analysis is performed using the scaling relations for the isotropic 3D Heisenberg model on the cubic lattice from Ref.~\cite{Hasenfratz1993} for lattices of size $L \times L \times L/2$ with $L$ up to 20 for $J_z/J_{xy}=0.05$ and $0.1$, and with $L\times L \times L$ lattices for a maximum $L$ of 12 for $J_z/J_{xy}=0.2-0.4$.

\begin{figure}
    \centering
    \includegraphics[width=\linewidth]{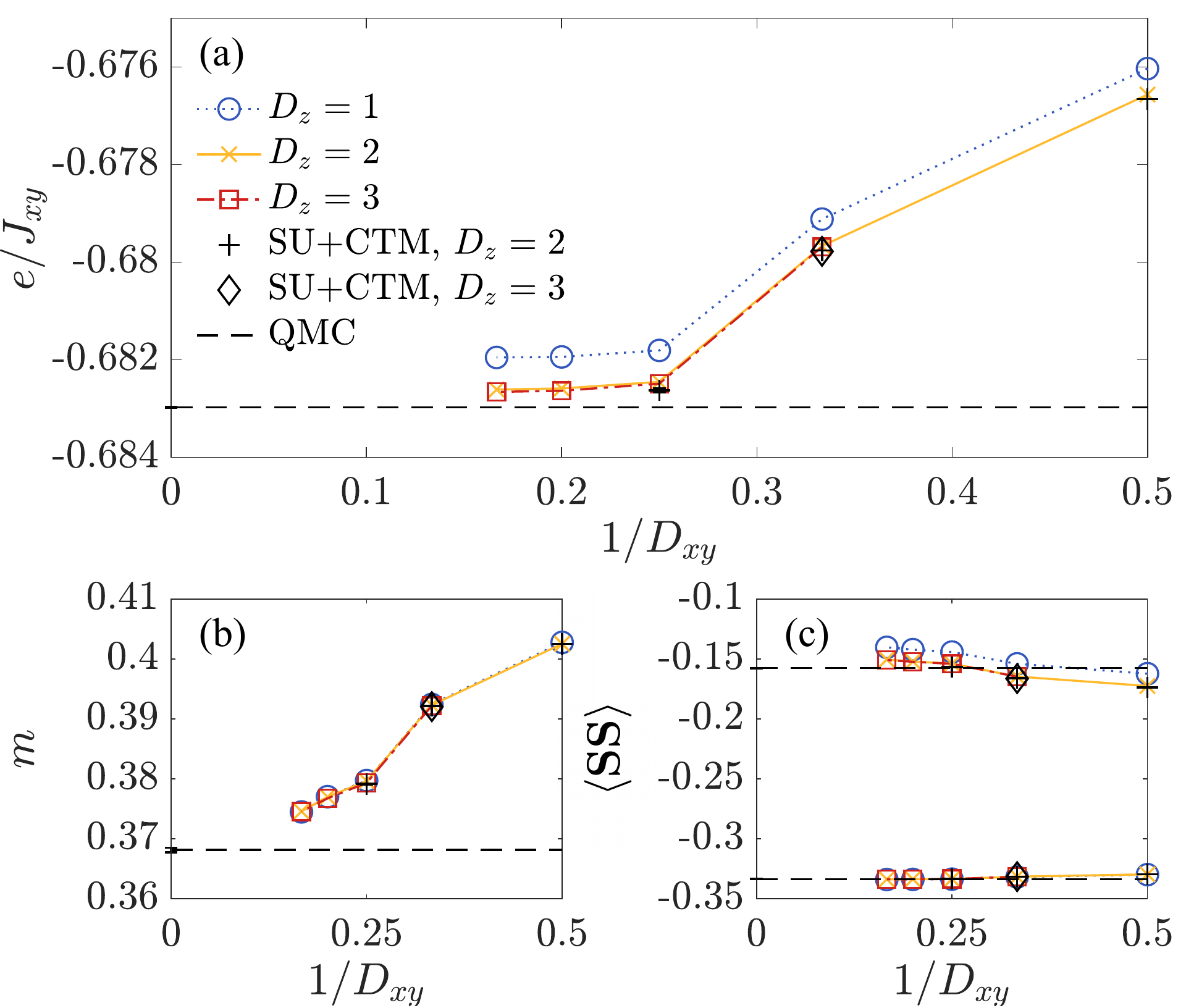}
    \caption{
    Results for the anisotropic 3D Heisenberg model with $J_z/J_{xy}=0.1$ as a function of  $1/D_{xy}$, obtained for $D_z=1-3$. For comparison, the extrapolated QMC results  are indicated by the horizontal dashed lines with the extrapolation error bar shown on the y-axis, which is of the order of the line width in (a) and (c). Data based on the full 3D contraction approach (SU+CTM) is shown by the black symbols.   (a)~Energy per site $e$ in units of $J_{xy}$. (b) Local magnetic moment $m$. (c) Nearest-neighbor spin-spin correlator in the z-direction, $\langle \mathbf{S} \mathbf{S} \rangle_z$ (top), and in the xy-direction, $\langle \mathbf{S} \mathbf{S} \rangle_{xy}$ (bottom).}
    \label{fig:aHeis_Jxy1_Jz01_layeredctm}
\end{figure}

We first consider the results for the energy per site, $e$,  for $J_z/J_{xy}=0.1$ in Fig.~\ref{fig:aHeis_Jxy1_Jz01_layeredctm}(a), plotted as a function of inverse bond dimension $D_{xy}$ for different values of $D_z$. 
Already a product of iPEPS layers  ($D_z=1$) yields a value that is remarkably close to the QMC result, with a relative error of only $0.15\%$ for $D_{xy}=6$. When $D_z$ is increased to 2, a significant improvement is found and the relative error at $D_{xy}=6$ is reduced to $0.05\%$, while a further increase to $D_z=3$ only yields a small enhancement. Overall, the improvement of the variational energy is clearly larger when increasing $D_{xy}$ (at least up to 4) compared to the improvement when increasing $D_z$, which further motivates the use of an anisotropic ansatz with $D_{xy}>D_z$. Comparing the LCTM scheme with the full 3D contraction (SU+CTM) only a small difference between the two methods is found.

Results for the local magnetic moment $m$ are shown in Fig.~\ref{fig:aHeis_Jxy1_Jz01_layeredctm}(b). Whereas $m$ systematically approaches the QMC result with increasing $D_{xy}$, the dependence on $D_z$ is small, suggesting that the reduction of the magnetic moment is predominantly due to the intraplane quantum fluctuations. 
The relative error of $m$ at $D_{xy}=6$ and $D_z=3$ is $1.7(1)\%$. In  Fig.~\ref{fig:aHeis_Jxy1_Jz01_layeredctm}(c) we present results for the nearest-neighbor spin-spin correlators in the intraplane and z-direction. The former is more accurately reproduced, which is a natural consequence of the fact that the latter enters with a prefactor $J_z/J_{xy}=0.1$ in the optimization of the tensors. Still, we find that the QMC result is approached with increasing $D_z$ at large $D_{xy}$ (note that increasing the two bond dimensions has an opposite effect on the change in the $\langle \mathbf{S} \mathbf{S} \rangle_z$ correlator~\cite{suppl_mat}).

\begin{figure}
    \centering
    \includegraphics[width=\linewidth]{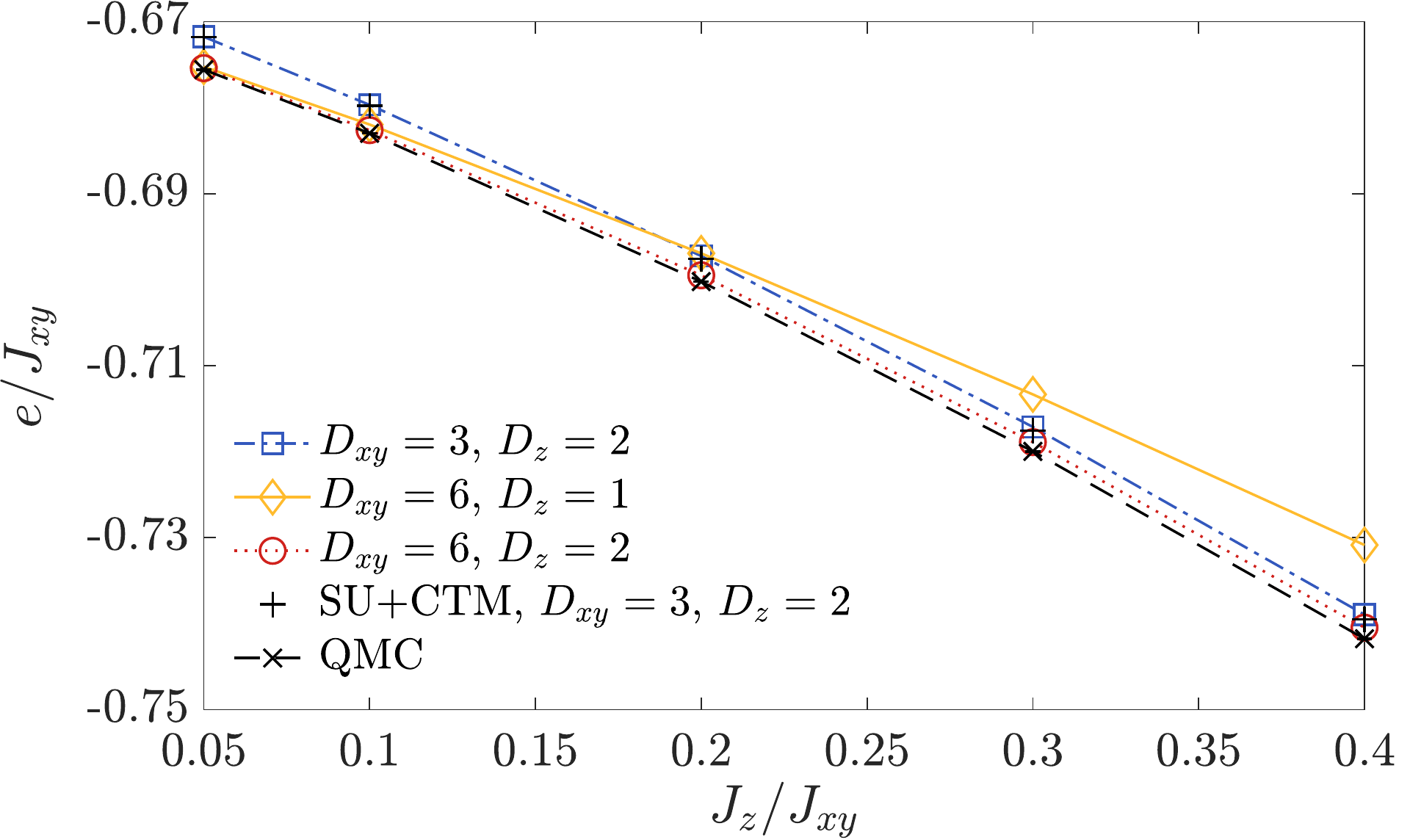}
    \caption{Energy per site as a function of $J_z/J_{xy}$ for different sets of bond dimensions, in comparison with data from a full 3D contraction (SU+CTM) and extrapolated QMC results.}
    \label{fig:aHeis_Jxy1_varJz_e}
\end{figure}

In Fig.~\ref{fig:aHeis_Jxy1_varJz_e} we present results for the energy per site as a function of $J_z/J_{xy}$, for selected values of $D_{xy}$ and $D_z$. Starting with the data for $D_{xy}=3$ and $D_z=2$ we find that the deviation with respect to the SU+CTM result slightly increases with increasing $J_z/J_{xy}$, although the deviation remains small even at a relatively large value of $J_z/J_{xy}=0.4$. For a small ratio $J_z/J_{xy}=0.05$, a product of iPEPS layers ($D_z=1$) for $D_{xy}=6$ already provides an energy close to the QMC result, with only a small improvement when increasing $D_z$ to 2. In contrast, for $J_z/J_{xy}=0.4$ the energy gain is large when increasing $D_z$, which is a natural consequence of the stronger entanglement between the layers for larger interlayer coupling.

We next consider a more challenging problem, the Shastry-Sutherland model~\cite{Shastry1981} (SSM) - a frustrated spin model relevant for SrCu$_2$(BO$_3$)$_2$~\cite{Miyahara2003} for which QMC suffers from the negative sign problem~\cite{wessel18}. It is described by square lattice layers of coupled dimers with Hamiltonian 
\begin{equation}
        \hat{H} = J \sum_{\langle i ,j \rangle} \mathbf{S}_i \mathbf{S}_j + J'\sum_{\langle i ,j \rangle'} \mathbf{S}_i \mathbf{S}_j + J'' \sum_{\langle i ,j \rangle''}  \mathbf{S}_i \mathbf{S}_j,
\end{equation}
with $J$ and $J'$ the intra- and interdimer couplings, $\mathbf{S}_i$ spin $S=1/2$ operators, and with an additional interlayer coupling $J''$ to the model as in Ref.~\cite{Koga2000} (see supplemental material~\cite{suppl_mat} for additional details).

\begin{figure}
    \centering
    \includegraphics[width=\linewidth]{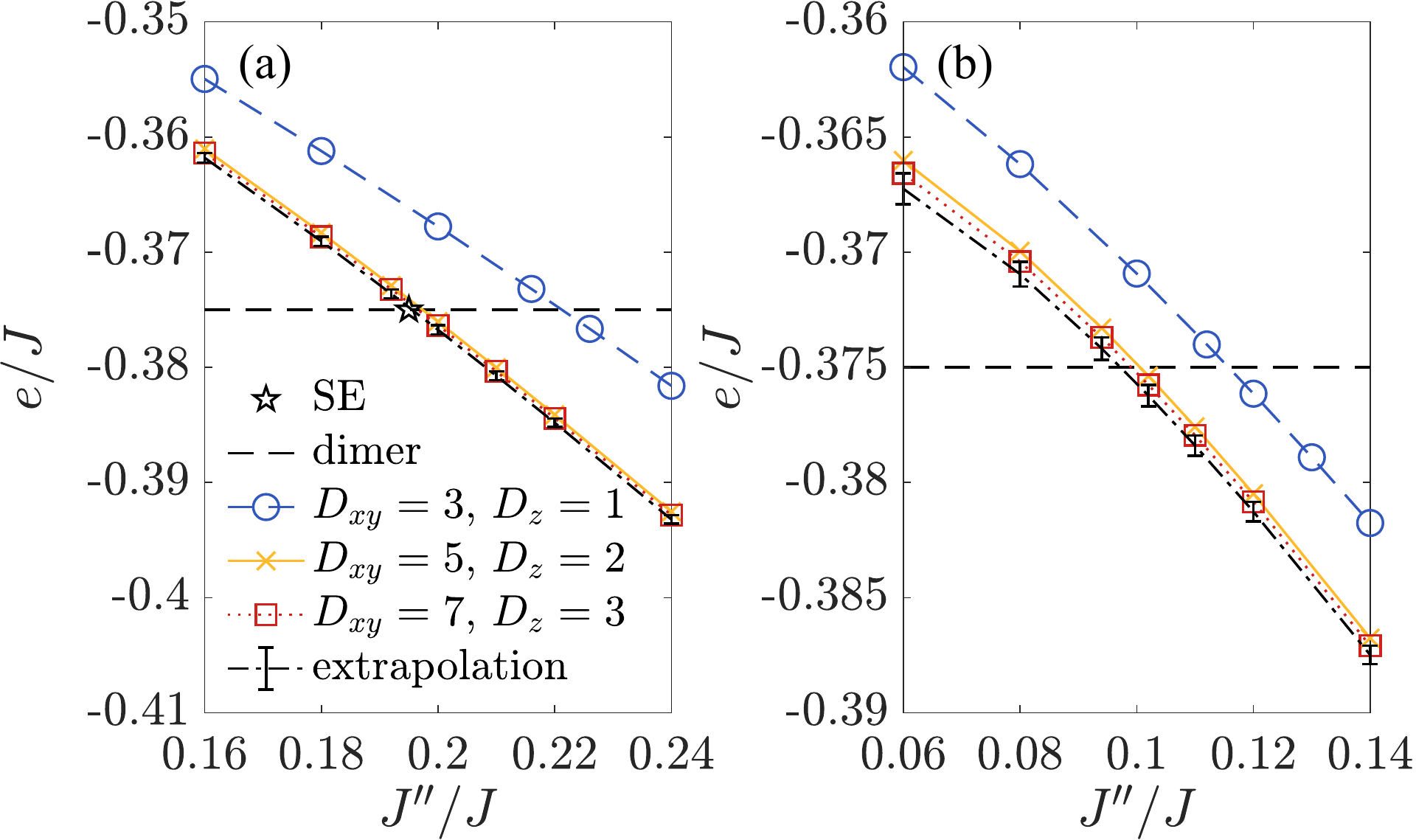}
    \caption{Energies per site as a function of $J''/J$ of the SSM, showing the phase transition between the exact dimer state (horizontal dashed line) and the antiferromagnetic state for different bond dimensions, for (a) $J'/J=0.61$ and (b) $J'/J=0.66$. The dashed-dotted line shows the energy extrapolated in inverse bond dimension. 
    }
      
    \label{fig:SSM_cut}
\end{figure}

We consider the dimer to antiferromagnetic phase transition for fixed $J'/J$ while varying $J''/J$ and compare it to fourth-order series expansion (SE) results~\cite{Koga2000}. 
The location of the phase transition is determined from the intersection of the exact energy of the dimer state ($-0.375J$ per site) with the energy of the antiferromagnetic state. For $J'/J=0.61$, shown in Fig.~\ref{fig:SSM_cut}(a), we find a close agreement with the SE result at large bond dimensions and also based on an extrapolation in inverse bond dimension~\cite{suppl_mat}. For larger values (within the dimer phase) no estimate from SE exists due to convergence problems~\cite{Koga2000}. With iPEPS, in contrast, we can  accurately determine the transition, as shown in Fig.~\ref{fig:SSM_cut}(b) for $J'/J=0.66$.  





\emph{Conclusions.} ---  
We have introduced the LCTM method which is an efficient approach to study layered 2D systems with a weak interlayer coupling. The main idea is to perform a decoupling of the 3D network using the FU truncation onto $D_z=1$ away from the center of each plane, while keeping the full bond dimension $D_z>1$ in the center, such that the resulting network can be efficiently contracted with the standard CTM method in each layer. 
Our benchmark results for the anisotropic Heisenberg model demonstrate that the method yields values in close agreement with a full 3D contraction (SU+CTM), at a substantially lower computational cost. The results are close to QMC results even for a small interlayer bond dimension $D_z=2$. Although the accuracy decreases when $J_z/J_{xy}$ is increased, errors remain relatively small up to $J_z/J_{xy}=0.4$. Our results for the SSM demonstrate that LCTM also enables the accurate study of problems which are out of reach of QMC due to the negative sign problem. 

There are several promising ways to further improve the LCTM method. First, the accuracy of the FU projection onto $D_z=1$ could be improved by making use of disentanglers between the layers~\cite{vidal2007-1,Evenbly2015}. Second, the accuracy of the contraction can be increased by including more $D_z>1$ bonds in the center~\cite{suppl_mat}, although at a higher computational cost. Instead of keeping $k$ open legs with total bond dimension $(D_z)^k$ in between the layers, the total dimension could be effectively reduced by introducing  appropriate projectors between the layers. Third, instead of a complete decoupling away from the center, a small vertical bond dimension $\chi_z$ could be kept in the CTM environment tensors in order to capture the most relevant interlayer entanglement away from the center. And finally, the contraction scheme may also be combined with an energy minimization based on automatic differentiation~\cite{liao2019} which is expected to provide more accurate tensors than the FFU optimization used here.

We believe our approach provides a powerful and practical tool for future studies of challenging layered 2D systems, especially models that are out of reach of QMC. Finally, we note that the LCTM method can be straightforwardly extended to fermionic systems and finite temperature calculations, e.g., by adapting ideas from Refs.~\cite{kraus2010,barthel2009,corboz2010,gu2010} and Refs.~\cite{kshetrimayum19,Czarnik2019}, respectively. 

\begin{acknowledgments}
This project has received funding from the European Research Council (ERC) under the European Union's Horizon 2020 research and innovation programme (grant agreement Nos. 677061 and 101001604). This work is part of the D-ITP consortium, a program of the Netherlands Organization for Scientific Research (NWO) that is funded by the Dutch Ministry of Education, Culture, and Science (OCW).
\end{acknowledgments}

\bibliography{references,refs}

\end{document}


\title{Supplemental Material for "Efficient tensor network algorithm for layered systems"}

\author{Patrick C.G. Vlaar}
\email[]{p.c.g.vlaar@uva.nl}
\affiliation{Institute for Theoretical Physics and Delta Institute for Theoretical Physics,
             University of Amsterdam, Science Park 904, 1098 XH Amsterdam, The Netherlands}

\author{Philippe Corboz}
\affiliation{Institute for Theoretical Physics and Delta Institute for Theoretical Physics,
             University of Amsterdam, Science Park 904, 1098 XH Amsterdam, The Netherlands}

\date{\today}

\maketitle

In this supplemental material, additional details and results on the LCTM method, the iPEPS optimization, and the Shastry-Sutherland model are presented. In Sec.~\ref{app:truncation}, the truncation used in the $D_z>1 \rightarrow D_z=1$ projection step and in the fast-full update (FFU) optimization is discussed as well as two computationally cheaper truncation approaches. Details on the FFU optimization and a comparison between FFU and other imaginary time evolution algorithms are presented in Sec.~\ref{app:optimization}. Section~\ref{app:singval_decay} discusses the decay of the singular value spectra of the iPEPS, which further justifies the use of an anisotropic ansatz. Additional results on the accuracy and convergence of the LCTM are provided in Sec.~\ref{app:accuracy}, including the dependence on the boundary bond dimension $\chi$, a comparison with the higher-order tensor renormalization group (HOTRG) method, the number of interlayer $D_z>1$ connections, and a comparison of different layer decoupling procedures. Finally, in Sec.~\ref{app:SSM} additional details on simulations of the Shastry-Sutherland model with interlayer coupling are given.

\section{Full update truncation} \label{app:truncation}
In this section, we provide more details on the full-update (FU) truncation method~\cite{Jordan2008,Phien2015}, which is used to perform the projection from $D_z>1$ to $D_z=1$ of the vertical bonds in the LCTM approach. The same scheme is also applied to truncate a bond within the FFU time evolution method (cf. next section). 

\begin{figure}
    \centering
    \includegraphics[width=\linewidth]{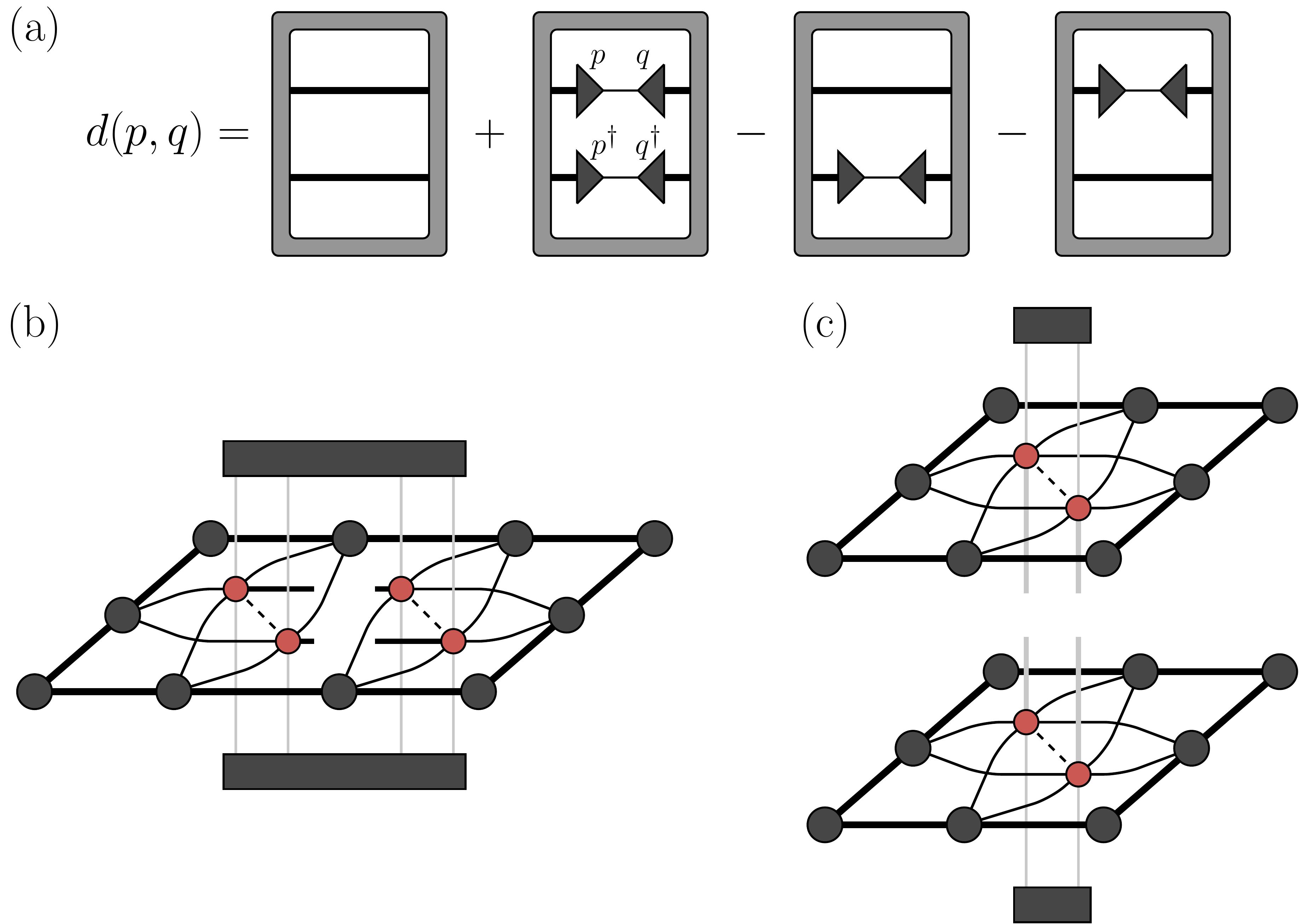}
    \caption{Relevant diagrams for the truncation scheme used in the LCTM method and in the FFU imaginary time evolution. (a) Diagrammatic representation of the norm distance in Eq.~\ref{eq:trunc_distance} which depends on the projectors $p$ and $q$ (triangles). The gray box represents the environment of the bond, and can be computed using the LCTM approach as shown in (b) and (c) for an intraplane and interplane bond, respectively. }
    \label{fig:fu_env}
\end{figure}

Let us consider an iPEPS $\ket{\psi}$ where we want to truncate a specific bond with bond dimension $\tilde D$ down to a smaller bond dimension $D$. We can do this by inserting a pair of projectors $(p,q)$ on the corresponding bond, yielding a truncated state $\ket{\psi'}$ which depends on $p$ and $q$. In order to find the optimal pair of projectors we need to minimize the norm distance between the two states,
\begin{align}
    d(p,q) &= \begin{Vmatrix} \ket{\psi} - \ket{\psi'} \end{Vmatrix}^2 \nonumber\\
    &= \langle{\psi}|\psi\rangle + \braket{\psi'}{\psi'} - \langle{\psi}|\psi'\rangle - \langle{\psi'}|\psi\rangle,  \label{eq:trunc_distance}
\end{align}
which is graphically represented  in Fig.~\ref{fig:fu_env}(a). The gray box corresponds to the environment of the bond to be truncated, i.e., the infinite tensor network surrounding the corresponding bond, which is the same for $\ket{\psi}$ and $\ket{\psi'}$ (since they differ only by the pair of projectors on the bond). The computation of these environments is done using the LCTM approach, and the corresponding diagrams for a horizontal and a vertical bond are shown in Figs.~\ref{fig:fu_env}(b) and Figs.~\ref{fig:fu_env}(c), respectively. 

We minimize Eq.~\ref{eq:trunc_distance} by an iterative scheme where we first keep $q$ fixed and minimize $d(p,q)$ with respect to $p$, which can be done by solving the resulting linear system~\cite{Phien2015}. Then $p$ is kept fixed and $d(p,q)$ is minimized with respect to $q$, and the procedure is repeated until convergence is reached. Alternatively, a conjugate gradient approach could also be used to minimize Eq.~\ref{eq:trunc_distance}. To improve the stability of the algorithm, the norm matrix is explicitly made Hermitian and positive definite as proposed in Ref.~\cite{Lubasch2014(2)}.

\subsection{Simple update and cluster update truncation} \label{app:truncation_sucu}
\begin{figure}
    \centering
    \includegraphics[width=\linewidth]{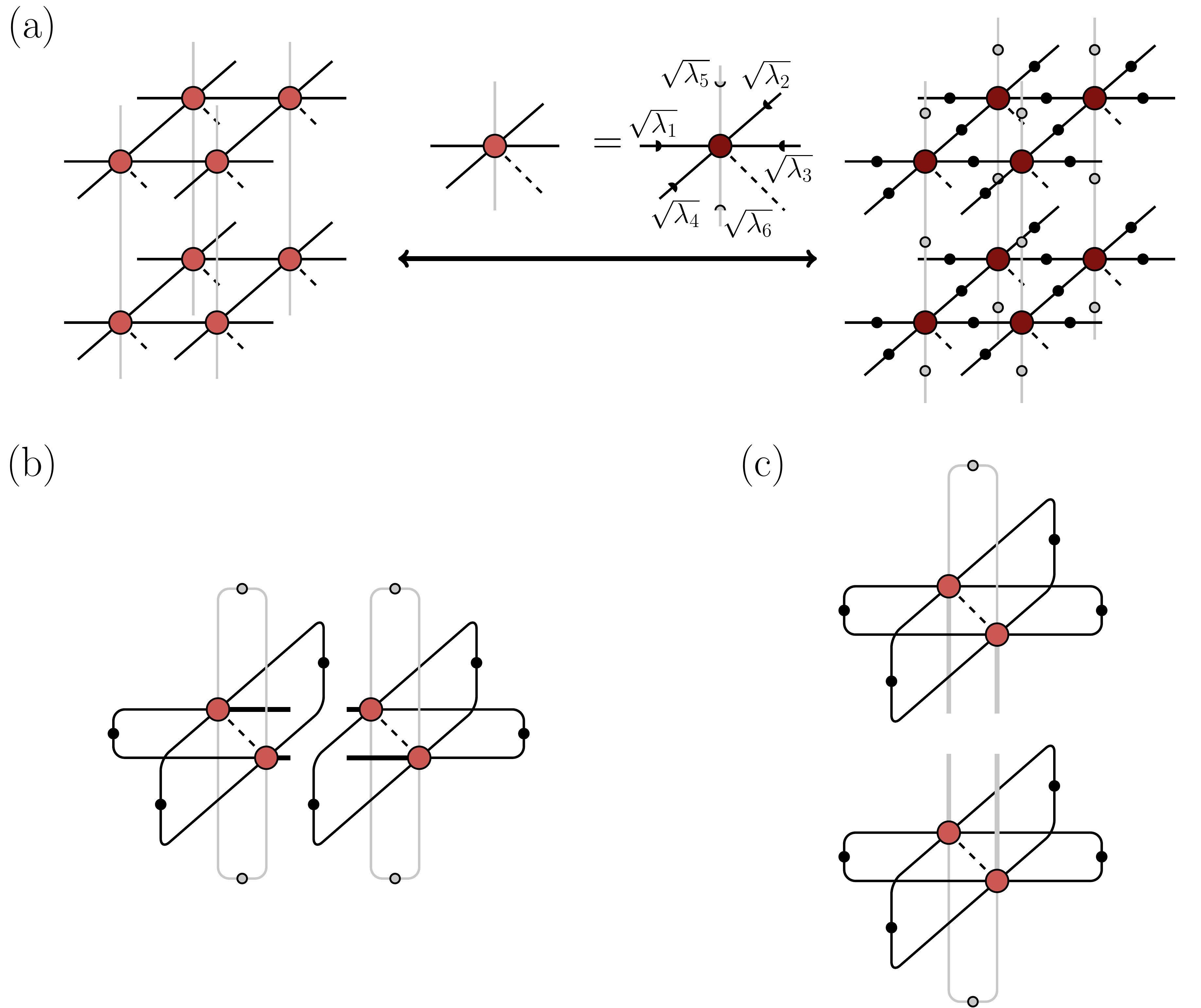}
    \caption{The environment used in the SU truncation. (a) An alternative formulation of the iPEPS ansatz in which singular value matrices $\lambda_i$ are introduced on the virtual bonds, represented by small circles. To retrieve the standard ansatz, the square root of the singular value matrices can be absorbed into the iPEPS tensors on each side. (b) The SU environment in the intraplane direction. (c) The SU environment in the interplane direction.
    }
    \label{fig:su_env}
\end{figure}

Besides the FU truncation, which takes the full wave function into account to truncate a bond, other schemes exist where the environment is approximated by only a finite number of tensors. These schemes are based on a modified iPEPS ansatz in which diagonal matrices with non-negative entries are introduced on each virtual bond~\cite{vidal2003-1,Jiang2008}, shown in Fig.~\ref{fig:su_env}(a). The matrices are naturally obtained during an imaginary time evolution algorithm as singular value matrix from an SVD of the projectors $p$ and $q$. Alternatively they can be extracted by an algorithm like in Ref.~\cite{Phien2015(2)}. The standard iPEPS tensors can be retrieved by absorbing the square root of the matrix on each bond into the tensor.  

In the simple update~(SU)~\cite{Jiang2008}, the truncation is performed based on an environment made of only two tensors together with their adjacent singular values, shown in Figs.~\ref{fig:su_env}(b) and (c) for intralayer and interlayer bonds respectively. 
Because the sites that are optimized are not connected through the environment, the truncation can also be directly performed by an SVD of the tensors including the singular values~\cite{Jiang2008}. The SU is computationally considerably cheaper than the FU, but it is also less accurate.

\begin{figure}
    \centering
    \includegraphics[width=\linewidth]{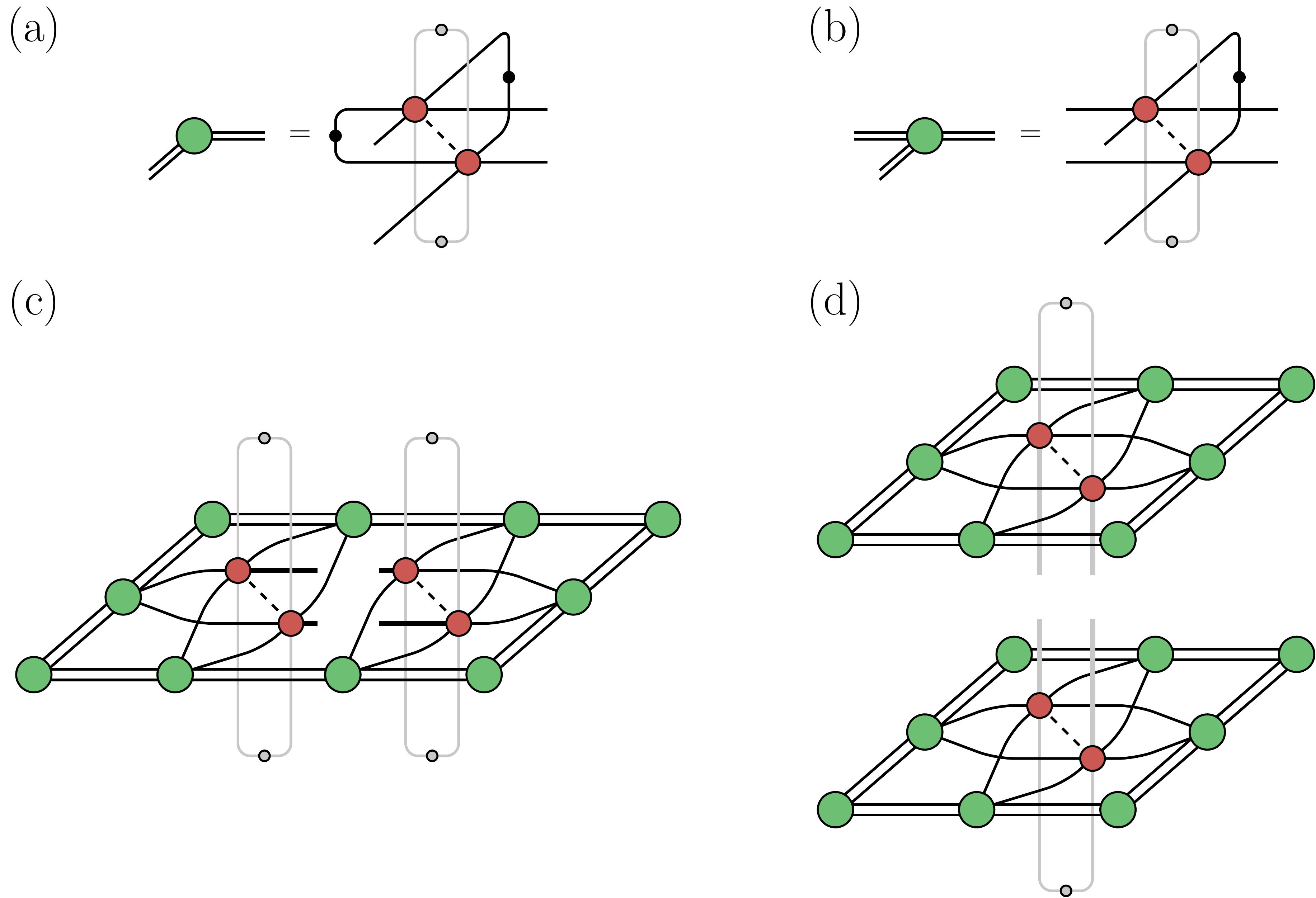}
    \caption{The environments used in our implementation of the CU truncation. (a+b) For graphical brevity, a compact notation is used. The small circles on the traced-out bonds represent the corresponding singular value matrices on the bonds. (c) A $4\times3$ cluster environment used for the intraplane update steps. (d) A $3\times3$ environment used in the interplane direction.}
    \label{fig:cu34_env}
\end{figure}

An improvement over the SU truncation (but still less accurate than the FU) is provided by the cluster update (CU) truncation~\cite{Wang2011,Lubasch2014,Lubasch2014(2),Dziarmaga2021}, which takes a larger environment consisting of a finite number of tensors around the bond into account to minimize Eq.~\ref{eq:trunc_distance}. Figure~\ref{fig:cu34_env}(c) and \ref{fig:cu34_env}(d) show examples of a $4\times3$ cluster environment for intraplane update steps and a $3\times3$ environment for the interplane bond respectively.

\section{Imaginary time evolution with LCTM} \label{app:optimization}

\begin{figure}
    \centering
    \includegraphics[width=\linewidth]{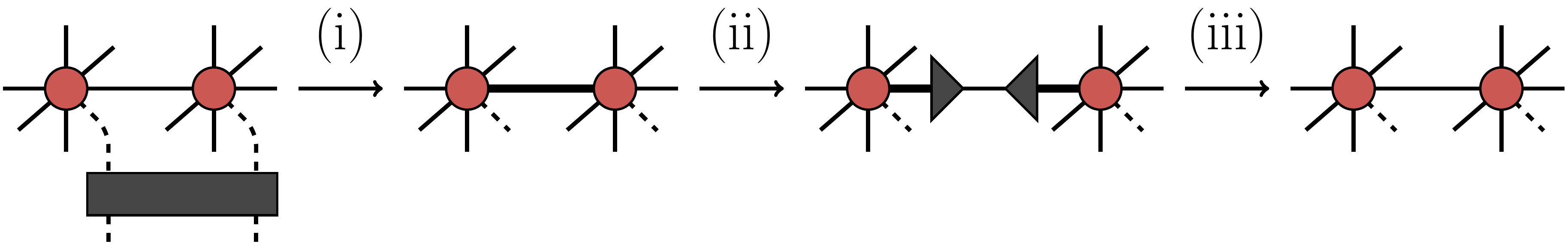}
    \caption{Application of an imaginary time evolution gate to the iPEPS ansatz. (i) Applying the gate increases the dimension of the respective bond to $\tilde D>D$. (ii) In order to avoid the bond dimension from growing exponentially, projectors are inserted which truncate the bond $\tilde D \rightarrow D$. (iii) Absorbing the projectors into the tensors provides the updated tensors.}
    \label{fig:time_gate}
\end{figure}

Imaginary time evolution is a well-established approach to obtain an iPEPS representing the ground state of a given Hamiltonian $\hat H$. The main idea is to project an initial state onto the ground state by applying the imaginary time evolution operator $e^{-\beta\hat{H}}$ with $\beta \rightarrow \infty$ on the initial state. 
For simplicity, let us consider a Hamiltonian consisting of only nearest-neighbor terms, $\hat{H}=\sum_i \hat{H}_i$. By using a Trotter-Suzuki decomposition the imaginary time evolution operator can be split into a product of local nearest-neighbor gates
\begin{equation}
    e^{-\beta \hat{H}} = (e^{-\tau \sum_i \hat{H}_i})^M = \prod_{j}^{M}\prod_i e^{-\tau \hat{H}_i} + \mathcal{O}(\tau),
\end{equation}
with $\beta = \tau M$. The error can be further reduced to $\mathcal{O}(\tau^2)$ using a second-order Trotter-Suzuki decomposition which is used in this work. The individual gates are sequentially applied to the iPEPS ansatz. Each application of a gate increases the bond dimension between the two sites to $\tilde D>D$, which needs to be truncated to avoid exponential growth of the bond dimension, as diagrammatically shown in Fig.~\ref{fig:time_gate}.

In the full update~(FU)~\cite{Jordan2008,Phien2015} method, the truncation is performed using the FU truncation discussed in Sec.~\ref{app:truncation}. The computational cost of this procedure is dominated by the environment contraction which must be computed every time a new gate is applied. The cost can be reduced with the fast-full update (FFU)~\cite{Phien2015}, which is the method used here. In the FFU, instead of performing a full convergence of the environment with the LCTM method after applying a gate, the environment from the previous time step is recycled, and, for intraplane bonds, only one single CTM iteration in the direction of the updated bond is performed, which is computationally cheaper. For bonds in the z-direction, no iteration of the CTM is required. The FFU is motivated by the observation that an application of a single time-evolution gate with a small time step has typically only a small effect on the environment. In addition to the CTM environment tensors, we also recycle both the projectors that truncate the iPEPS tensors to $D_z\rightarrow1$ as well as the dominant eigenvectors carrying contributions from the other layers that are used in the x- and y-update steps. Both are recomputed after the updates in z-direction. 

\begin{figure}
    \centering
    \includegraphics[width=\linewidth]{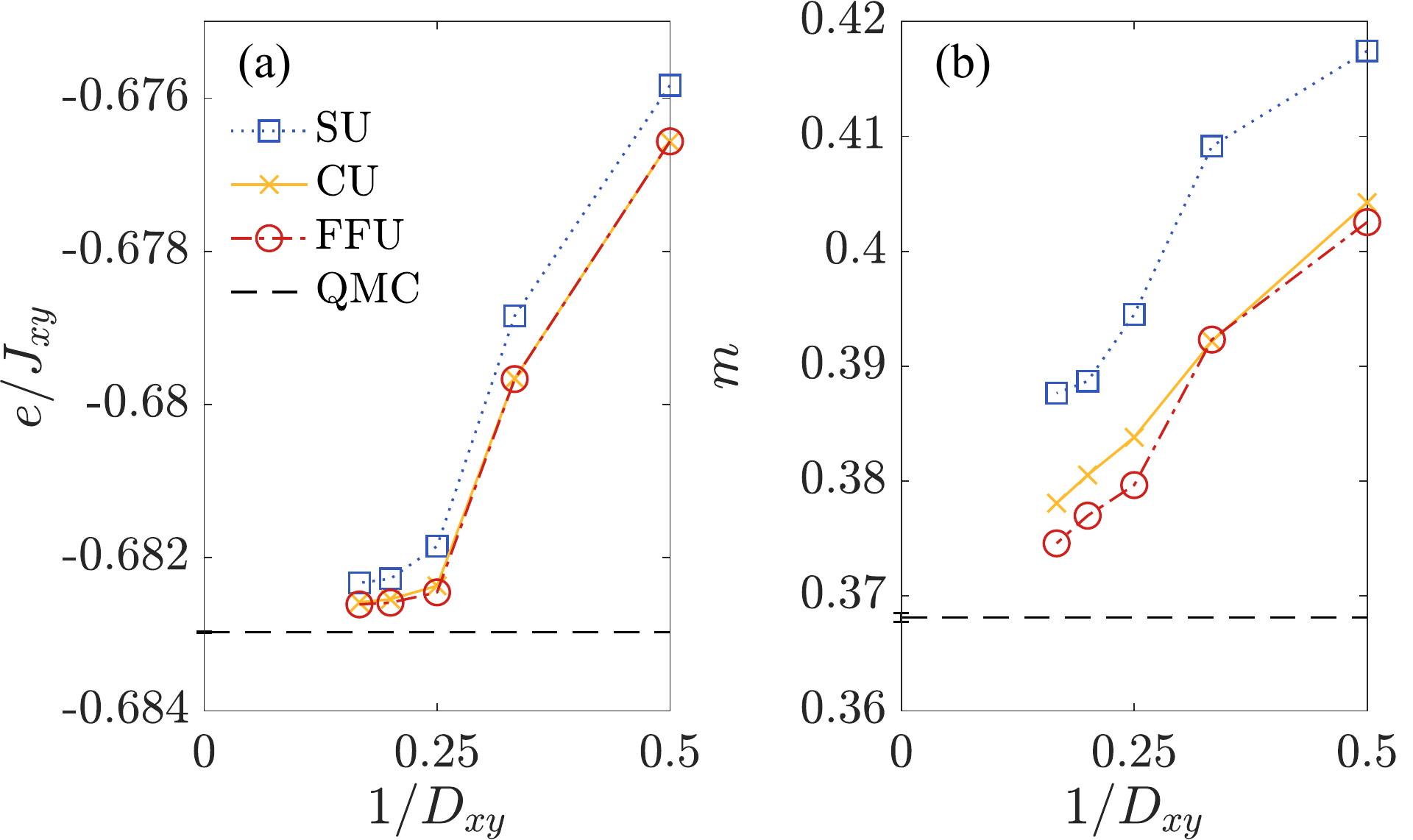}
    \caption{Comparison of results obtained with the SU, CU, and FFU imaginary time evolution methods for the anisotropic Heisenberg model with $J_z/J_{xy}=0.1$ as a function of $1/D_{xy}$ and $D_z=2$. For reference, extrapolated quantum Monte Carlo (QMC) results are provided. (a) Energy per site $e$ in units of $J_{xy}$. (b) Local magnetic moment $m$.}
    \label{fig:aHeis_Jxy1_Jz01_update}
\end{figure}

In the main text, results have been obtained based on the FFU approach. Alternatively, we have also tested imaginary time evolution schemes based on a SU and CU truncation, which are computationally cheaper but also less accurate.  
Here, we compare results of these variants obtained for the anisotropic Heisenberg model with $J_z/J_{xy}=0.1$ in Fig.~\ref{fig:aHeis_Jxy1_Jz01_update}.
The SU leads to slightly higher energies than the FFU, as shown in Fig.~\ref{fig:aHeis_Jxy1_Jz01_update}(a), whereas the CU yields similar results to the FFU. For the local magnetic moment $m$, shown in Fig.~\ref{fig:aHeis_Jxy1_Jz01_update}(b), the deviation between the SU and FFU is more pronounced. Also here, the CU gives a significant improvement upon the SU result, with slightly larger values compared to the FFU result at large bond dimensions.



\section{Decay singular value spectrum} \label{app:singval_decay}
\begin{figure}
    \centering
    \includegraphics[width=\linewidth]{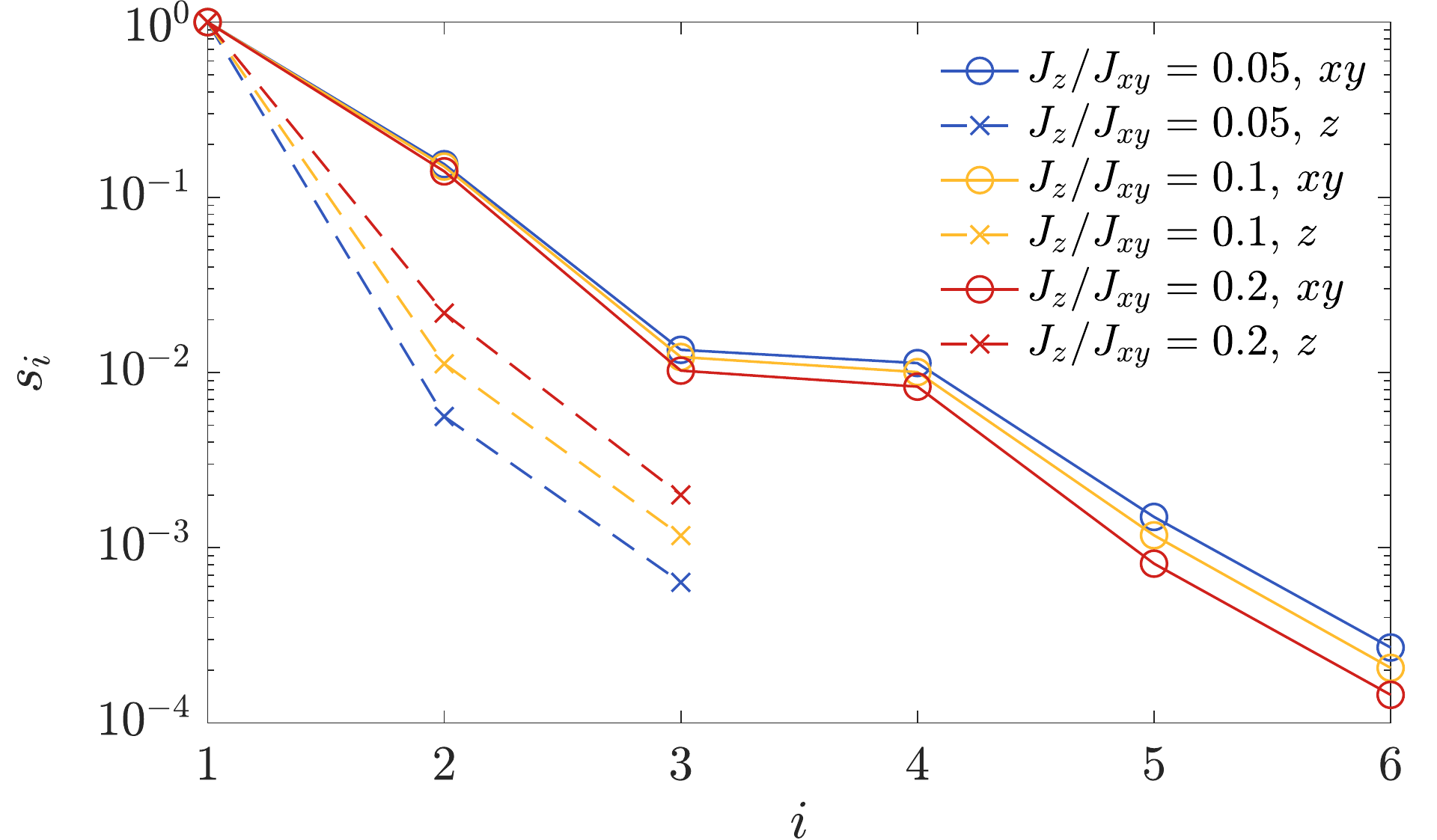}
    \caption{Spectrum of the singular values $s_i$ on the intra- and interplane bonds in the iPEPS ansatz obtained for the anisotropic Heisenberg model with $D_{xy}=6$ and $D_z=3$, exhibiting a fast decay (weaker entanglement) in the z-direction and a slow decay (stronger entanglement) in the xy-direction for small $J_z/J_{xy}$.}
    \label{fig:aHeis_Jxy1_Jz01_singvals}
\end{figure}

To further motivate the anisotropic ansatz and contraction approach, we consider the singular value spectrum on the intra- and interplane bonds in Fig.~\ref{fig:aHeis_Jxy1_Jz01_singvals} obtained for the 3D anisotropic Heisenberg model. 
The singular value matrices are extracted from our FFU-optimized tensors using the algorithm from Ref.~\cite{Phien2015(2)}. 

A much faster decay of the singular values can be observed in the z-direction than in the intraplane direction, as expected, due to the weak entanglement between the planes. Increasing $J_z/J_{xy}$ leads to a slower decay in the z-direction, suggesting that the value of $D_z$ needs to be increased. Eventually, for sufficiently large $J_z/J_{xy}$ the singular values in all directions will become of similar magnitude, such that an anisotropic ansatz in combination with the LCTM contraction is no longer justified.

\section{Accuracy of the LCTM contraction} \label{app:accuracy}

The accuracy of the LCTM contraction is controlled by both the boundary bond dimension $\chi$ of the CTM environment tensors as well as the number of untruncated interlayer connections that are kept in the center of the network. In this section, we analyze the dependence of the results on these parameters for the 3D anisotropic Heisenberg model. We also provide a comparison to the higher-order tensor renormalization group (HOTRG) method~\cite{xie12,Iino2019}. Finally, we examine alternative approaches for the $D_z>1$ to $D_z=1$ truncation performed away from the center. 

\subsection{Convergence in \texorpdfstring{$\chi$}{χ}} \label{app:convergence_ctm}

\begin{figure}
    \centering
    \includegraphics[width=\linewidth]{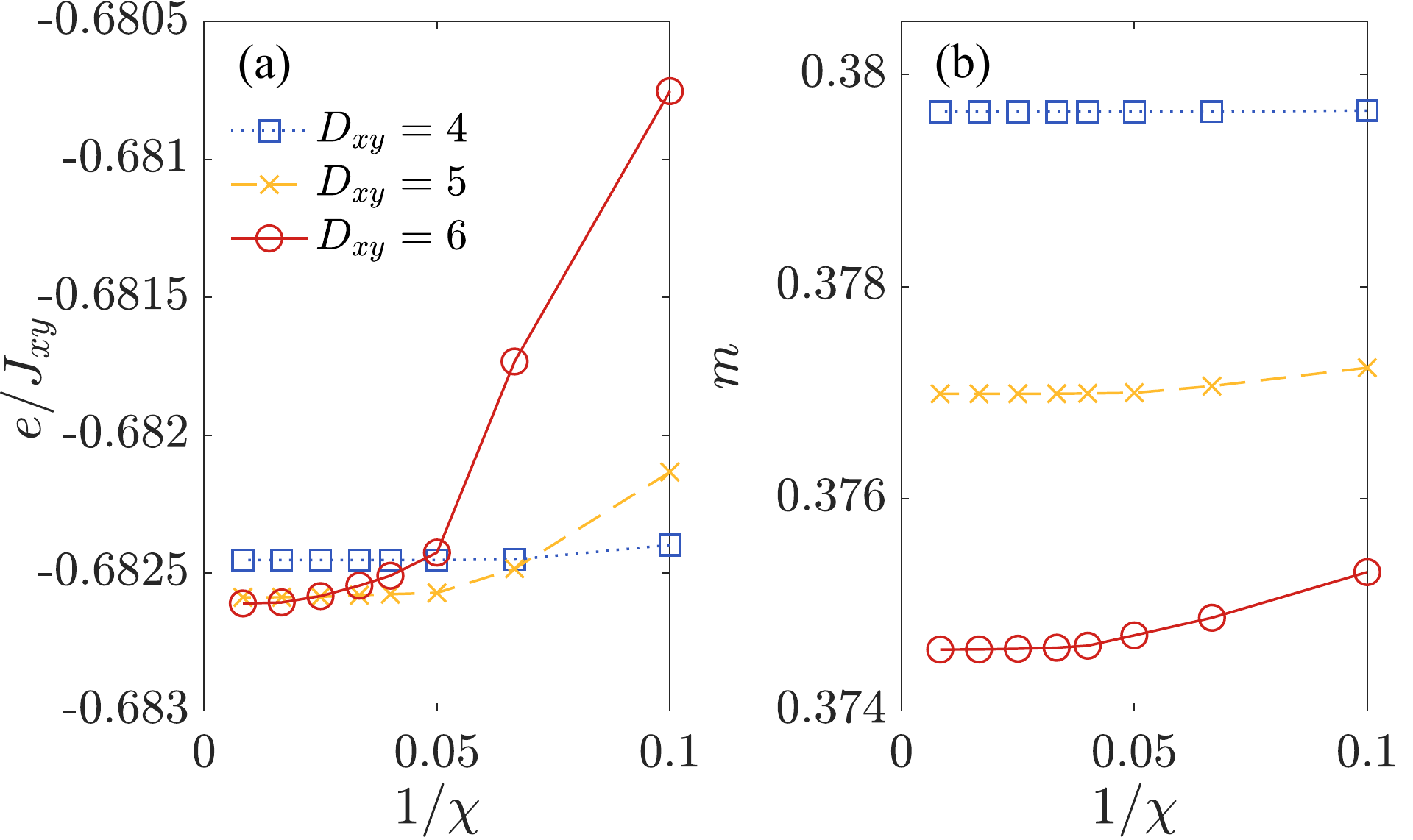}
    \caption{Convergence as a function of the inverse CTM boundary dimension $1/\chi$ for $D_{xy}=4-6$ and $D_z=2$. The results are for the anisotropic Heisenberg model with $J_z/J_{xy}=0.1$. (a) The energy per site $e$ in units of $J_{xy}$. (b) Local magnetic moment $m$.}
    \label{fig:aHeis_Jxy1_Jz01_chic}
\end{figure}

In Fig.~\ref{fig:aHeis_Jxy1_Jz01_chic}(a) we show the energy per site, $e$, for $J_z/J_{xy}=0.1$ as a function of $1/\chi$ for different values of $D_{xy}$ and $D_z=2$. Convergence is reached at sufficiently large $\chi$, where a higher $\chi$ is needed for larger $D_{xy}$ for an accurate evaluation, as expected. Interestingly, we find that an increase in $D_z$ does not require a larger $\chi$ to converge. Similar observations can be made for the local magnetic moment $m$ in Fig.~\ref{fig:aHeis_Jxy1_Jz01_chic}(b). For the results in the main text, $\chi$ is chosen sufficiently large such that  finite-$\chi$ errors are negligible.

\subsection{Comparison with HOTRG} \label{app:HOTRG}
In this section we present a comparison of the convergence behavior of the LCTM method as a function of $\chi$ with results obtained with the higher-order tensor renormalization group (HOTRG) approach~\cite{xie12,Iino2019}, which is another method to contract 3D tensor networks. Figure~\ref{fig:aHeis_Jxy1_Jz01_chic} shows the results for the energy and magnetization for different bond dimensions and $D_z=2$, using tensors obtained from a FFU optimization (the same optimized tensors are used for the two contraction approaches). 
For small bond dimension $D_{xy}=2$ the two approaches yield similar results at large $\chi$. However, with increasing bond dimension it becomes more and more challenging to reach convergence with HOTRG, which overall displays an irregular convergence behavior as a function of $\chi$  and which is computationally substantially more expensive than LCTM. Similar observations have been made for the isotropic 3D Heisenberg model in Ref.~\cite{Vlaar2021}, in which it was found that SU+CTM shows a much more regular convergence behavior than HOTRG. For this reason we have taken results from SU+CTM as reference values in the main text.

\begin{figure}
    \centering
    \includegraphics[width=\linewidth]{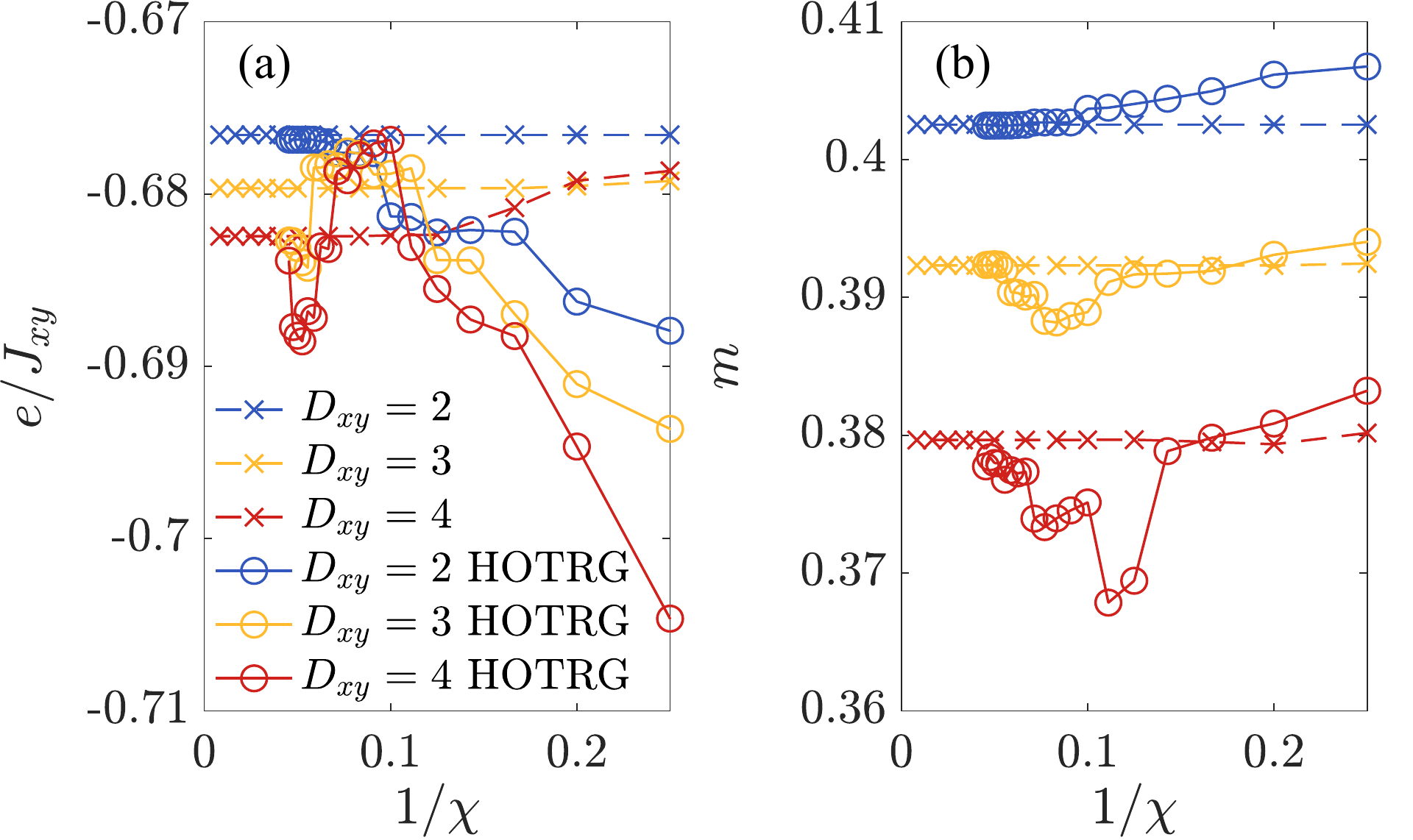}
    \caption{Comparison between LCTM and HOTRG for the convergence of the energy (a) and the magnetization (b) as a function of inverse bond dimension $1/\chi$.  }
    \label{fig:aHeis_Jz01_Dz2_hotrg}
\end{figure}

\subsection{Interlayer connectivity}\label{app:accuracy_star}
\begin{figure}
    \centering
    \includegraphics[width=\linewidth]{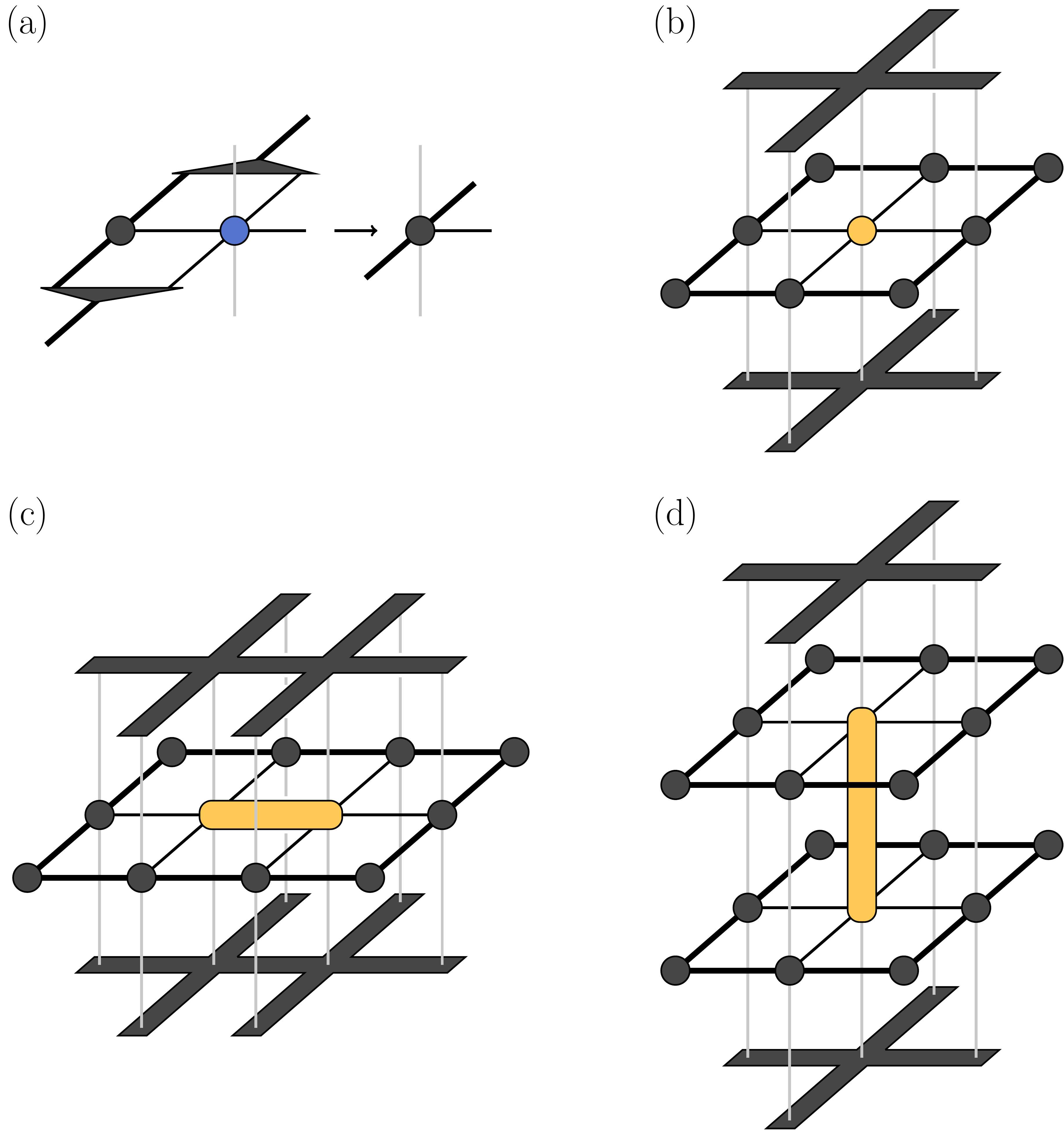}
    \caption{The star LCTM contraction which differs from the standard LCTM because both the central and nearest-neighbor interlayer connections remain untruncated. The yellow tensors indicate the contraction of bra- and ket-tensors with an operator in between. (a) The star LCTM differs from the standard LCTM at the final CTM step where a tensor with untruncated interlayer connections is absorbed into the environment. (b) The final diagram for the computation of a one-site observable. The leading eigenvector is a larger object compared to the standard LCTM due to the additional untruncated bonds. (c) The diagram of an intraplane two-site observable. (d) The diagram of an interplane two-site observable.} 
    \label{fig:layered_ctm_star}
\end{figure}

The number of untruncated $D_z>1$ interlayer connections is another parameter controlling the accuracy of the LCTM method. The results in the main text have been obtained by just keeping a single connection in the center for one-site observables and interlayer two-site observables, and two connections for intralayer two-site observables. Here we present a comparison to a different scheme, in which we also keep the interlayer connections on the tensors neighboring the (two) central one(s). In practice, this can be implemented by absorbing tensors with $D_z>1$ into the environment tensors at the final CTM step, as depicted in Fig.~\ref{fig:layered_ctm_star}(a). Figures~\ref{fig:layered_ctm_star}(b)-(d) show the diagrams to evaluate a one-site observable, an intraplane two-site observable, and an interplane two-site observable, respectively. We call this scheme the star LCTM. It has the advantage that more of the interlayer correlations are taken into account, however, at the expense of a significantly higher contraction cost.

\begin{figure}
    \centering
    \includegraphics[width=\linewidth]{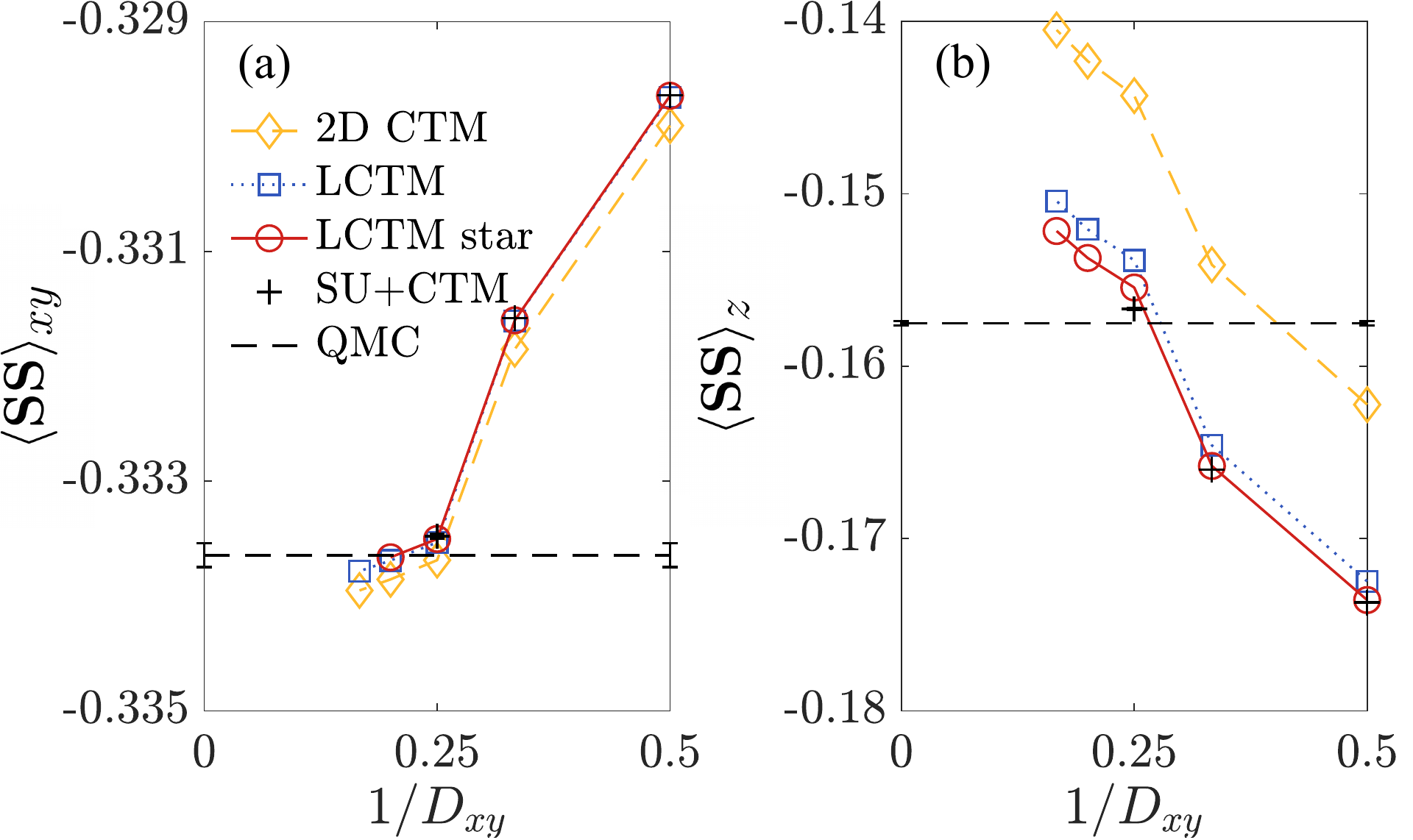}
    \caption{Contraction results for the nearest-neighbor spin-spin correlators obtained with a 2D CTM, standard LCTM, and star LCTM for the anisotropic Heisenberg model with $J_z/J_{xy}=0.1$ as a function of $1/D_{xy}$ and with $D_z=2$. For comparison, results obtained from a full 3D contraction (SU+CTM) and extrapolated QMC results are provided. (a)~Nearest-neighbor spin-spin correlator in the intraplane direction $\langle \mathbf{S} \mathbf{S} \rangle_{xy}$. (b) Nearest-neighbor spin-spin correlator in the interplane direction $\langle \mathbf{S} \mathbf{S} \rangle_z$.}
    \label{fig:aHeis_Jxy1_Jz01_23star_ss}
\end{figure}

In Fig.~\ref{fig:aHeis_Jxy1_Jz01_23star_ss}(a) results for the nearest-neighbor spin-spin correlator in the intraplane direction are shown, using different contraction schemes to evaluate them. The tensors have been obtained for $J_z/J_{xy}=0.1$ using the FFU imaginary time evolution based on the standard LCTM scheme. Besides the standard and star LCTM approach, we also include data from the 2D CTM in which no interlayer connections are kept (i.e. also the connection on the central tensor is truncated to $D_z=1$), and from the full 3D contraction (SU+CTM) which we take as reference values. We observe that without the interlayer connections (2D CTM) the deviation from the SU+CTM result is relatively large, whereas  both the standard and star LCTM show a close agreement with the full 3D contraction.

Figure~\ref{fig:aHeis_Jxy1_Jz01_23star_ss}(b) shows results in the interlayer direction. The deviation from the SU+CTM result is larger here with the standard LCTM scheme, although it performs much better than the 2D CTM. A significant improvement is obtained by using the star LCTM, with a close agreement to SU+CTM for $D_{xy}=2$ and $3$. For larger $D_{xy}$ we expect that the results can be further improved by keeping even more interlayer connections.

As already pointed out in the main text, increasing the two bond dimensions $D_{xy}$ and $D_z$ has an opposite effect on the change in the $\langle \mathbf{S} \mathbf{S} \rangle_z$ correlator, which can be intuitively understood as follows. Firstly, an increase in $D_z$ at fixed $D_{xy}$ naturally leads to a decrease in $\langle \mathbf{S} \mathbf{S}  \rangle_z$, because the higher bond dimension lowers the variational energy on these bonds. The increase in $\langle \mathbf{S} \mathbf{S}  \rangle_z$ with increasing $D_{xy}$ is less obvious, but can be best understood in the $D_z=1$ limit. In this limit the $\langle \mathbf{S} \mathbf{S}  \rangle_z$ correlator (or equivalently the energy on a z-bond) is minimized by the classical antiferromagnetic state which is realized for $D_{xy}=1$, and amounts to $-m^2$, with $m=1/2$ the magnitude of the local magnetic moment on each site. By increasing $D_{xy}$, $m$ will become smaller due to intraplane quantum fluctuations (entanglement), and hence the $\langle \mathbf{S} \mathbf{S} \rangle_z$ correlator  between the layers with $D_z=1$ will increase. We note that the opposite effect is observed for $\langle \mathbf{S} \mathbf{S} \rangle_{xy}$, just that the dependence on $D_z$ is very weak.

\subsection{Alternative layer decoupling approaches} \label{app:accuracy_decoupling}

A key step in the LCTM method is the decoupling of the layers away from the center, which for the results in the main text is done by a FU truncation to $D_z=1$. In this section, we compare several alternative local truncation approaches to the FU results.

The first alternative we consider is to trace out the bonds in the z-direction by connecting the respective bonds of the iPEPS tensors in the bra- and in the ket-layers. On these bonds, we include the corresponding singular value matrices in the same spirit as done in the SU and CU truncation. Another option we test is the SU truncation. Finally, we consider a CU truncation based on the $3\times3$ environment shown in Fig.~\ref{fig:cu34_env}(d).

\begin{figure}
    \centering
    \includegraphics[width=\linewidth]{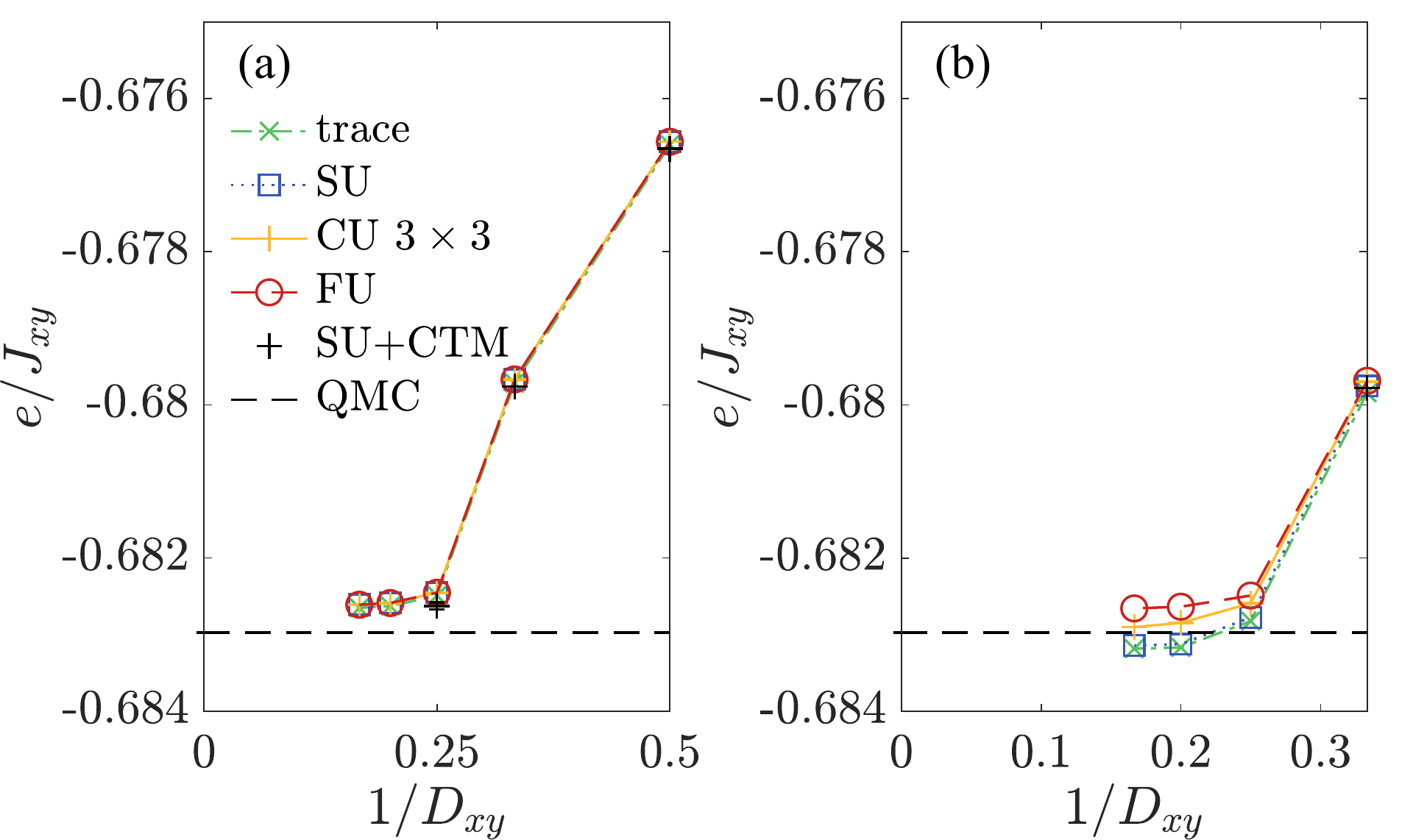}
    \caption{Results obtained with alternative layer decoupling approaches based on tracing out the z-bond, a SU truncation, and a CU truncation using a $3\times3$ cluster size compared to a FU truncation. The energy per site is shown as a function of $1/D_{xy}$ for the anisotropic Heisenberg model with $J_z/J_{xy}=0.1$. For reference, a SU+CTM contraction and an extrapolated QMC result are provided as well. (a) $D_z=2$. (b) $D_z=3$.}
    \label{fig:aHeis_Jxy1_Jz01_Dz1trunc}
\end{figure}

Figure~\ref{fig:aHeis_Jxy1_Jz01_Dz1trunc}(a) shows results for the energy per site, $e$, as a function of $1/D_{xy}$ and $D_z=2$ at $J_z/J_{xy}=0.1$. Here we find that the alternative truncation approaches show a good agreement with the FU layer decoupling scheme. For the $D_z=3$ case, presented in Fig.~\ref{fig:aHeis_Jxy1_Jz01_Dz1trunc}(b), however, the scheme based on tracing out the interlayer connections and the SU truncation both give a significant underestimation of the energy compared to FU. Although the CU $3\times3$ truncation performs better, it yields values that are too small as well. These results indicate that performing an accurate truncation is important, at least for $D_z>2$, and they motivate the use of the computationally more expensive FU truncation in the main text.

\section{Details on the simulations of the Shastry-Sutherland model with interlayer coupling} \label{app:SSM}
\begin{figure}
    \centering
    \includegraphics[width=\linewidth]{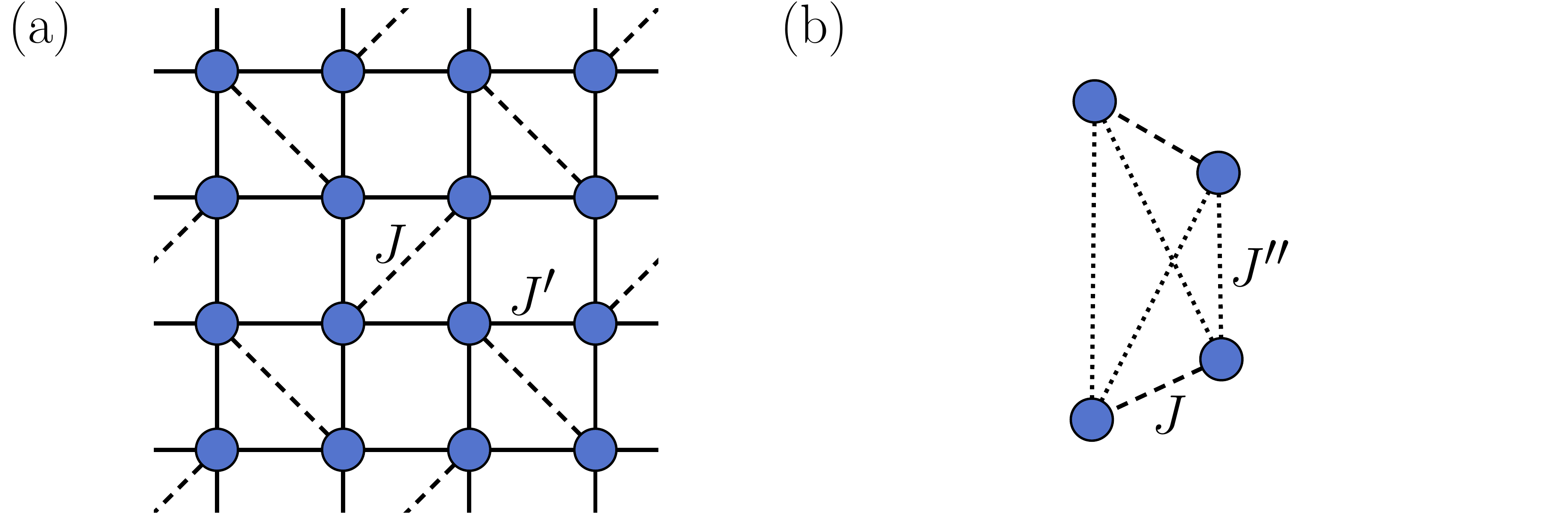}
    \caption{Lattice structure of the Shastry-Sutherland model with interlayer coupling. (a) The 2D SSM consists of orthogonal dimers with $J$ the intra- and $J'$ the interdimer coupling  indicated by dashed and solid lines, respectively. (b) In the 3D model the layers are stacked on top of each other such that dimers in neighboring layers are orthogonal to each other. Each site of a dimer interacts with both sites of the neighboring dimer in the adjacent layer with an interlayer coupling  $J''$ shown by the dotted lines.}
    \label{fig:SSM_lattice}
\end{figure}

In this section we provide additional details on the iPEPS simulations of the Shastry-Sutherland model (SSM) with interlayer coupling. The 3D lattice structure with the intra- and interdimer couplings $J$ and $J'$, and interplane coupling $J''$, is presented in Fig.~\ref{fig:SSM_lattice}.

In our numerical simulations, a cubic iPEPS ansatz is used, with one tensor per dimer as previously done for the 2D model~\cite{Corboz2013}. To improve the efficiency we use tensors with $U(1)$ symmetry. To obtain the energies of the antiferromagnetic states across the phase transition, we initialize the FFU optimizations from states obtained in the antiferromagnetic phase. Thanks to hysteresis effects across the first order phase transition, the state  remains in the antiferromagnetic phase even beyond the critical coupling, which enables us to accurately determine the phase transition by the intersection of the energy with the dimer energy (which is exactly $-0.375J$ per site). 

The extrapolation of the energy shown in Fig.~5 in the main text is done based on a linear extrapolation in $1/\kappa$, with $D_{xy}=\kappa$ and $D_z=(\kappa-1)/2$, using the data for ($D_{xy}=5$, $D_z=2$), ($D_{xy}=7$, $D_z=3$) as well as the mean value of ($D_{xy}=6$, $D_z=2$) and ($D_{xy}=6$ and $D_z=3$). Since the energy typically converges faster than linearly in $1/D$ we take the average of the extrapolated value and largest $D$ value as the infinite $D$ estimate, and its difference to the largest $D$ value as an estimate of the error bar.

\bibliography{references,refs}